\documentclass[letterpaper]{article}
\usepackage{isea}
\usepackage[pdftex]{graphicx}
\usepackage{times}
\usepackage{helvet}
\usepackage{hyperref}
\usepackage{courier}
\usepackage[numbers]{natbib}
\usepackage{orcidlink}

\title{Cybroc: Cyborgizing Broccoli for Longevity}

\author{Ke Huang\textsuperscript{1}, Yue Zhou\textsuperscript{2}, Xi He\textsuperscript{3},  Weibo Chen\textsuperscript{4},
Botao Amber Hu\textsuperscript{5}~\thanks{Corresponding author; also serves as a visiting lecturer at the School of Design \& Innovation, China Academy of Art.
} \\
\textsuperscript{1,2,3,4} School of Design \& Innovation, China Academy of Art; \textsuperscript{5}  Reality Design Lab\\
 Hangzhou, China;  New York City, USA\\
kehuang81@gmail.com,  fallonyueue@gmail.com, xiiiris625@gmail.com, chenweibo212@gmail.com,
botao@reality.design\\
}

\begin{document}
\maketitle

\begin{abstract}
\emph{Cybroc} is a series of kinetic art installations exploring the recent proliferating populist longevity activism through the satirical cyborgization of broccoli. The artwork augments the symbol of health food—broccoli—with prosthetic limbs to perform so-called longevity-enhancing exercises such as cold plunges, treadmill running, brachiation (arm-swinging), sled pushing, etc.—all simulations of primal human survival tasks reframed as modern fitness routines. Despite its mechanical augmentations, the broccoli's inevitable decay and rotting after exhibiting high-intensity performances prompts reflection on the limits of biological enhancement and the ethics of human enhancement beyond natural capabilities, particularly transhumanist ideals. By juxtaposing a symbolic healthy vegetable with cutting-edge concepts of human enhancement, \emph{Cybroc} challenges viewers to consider the intersection of nature, technology, and the human quest for extended lifespan in our transhuman era.
\end{abstract}

{\small Transhumanism, Longevity Activism,
Cyborgization, Kinetic Art, Morphological Freedom}

\begin{figure}
    \centering
    \includegraphics[width=1\linewidth]{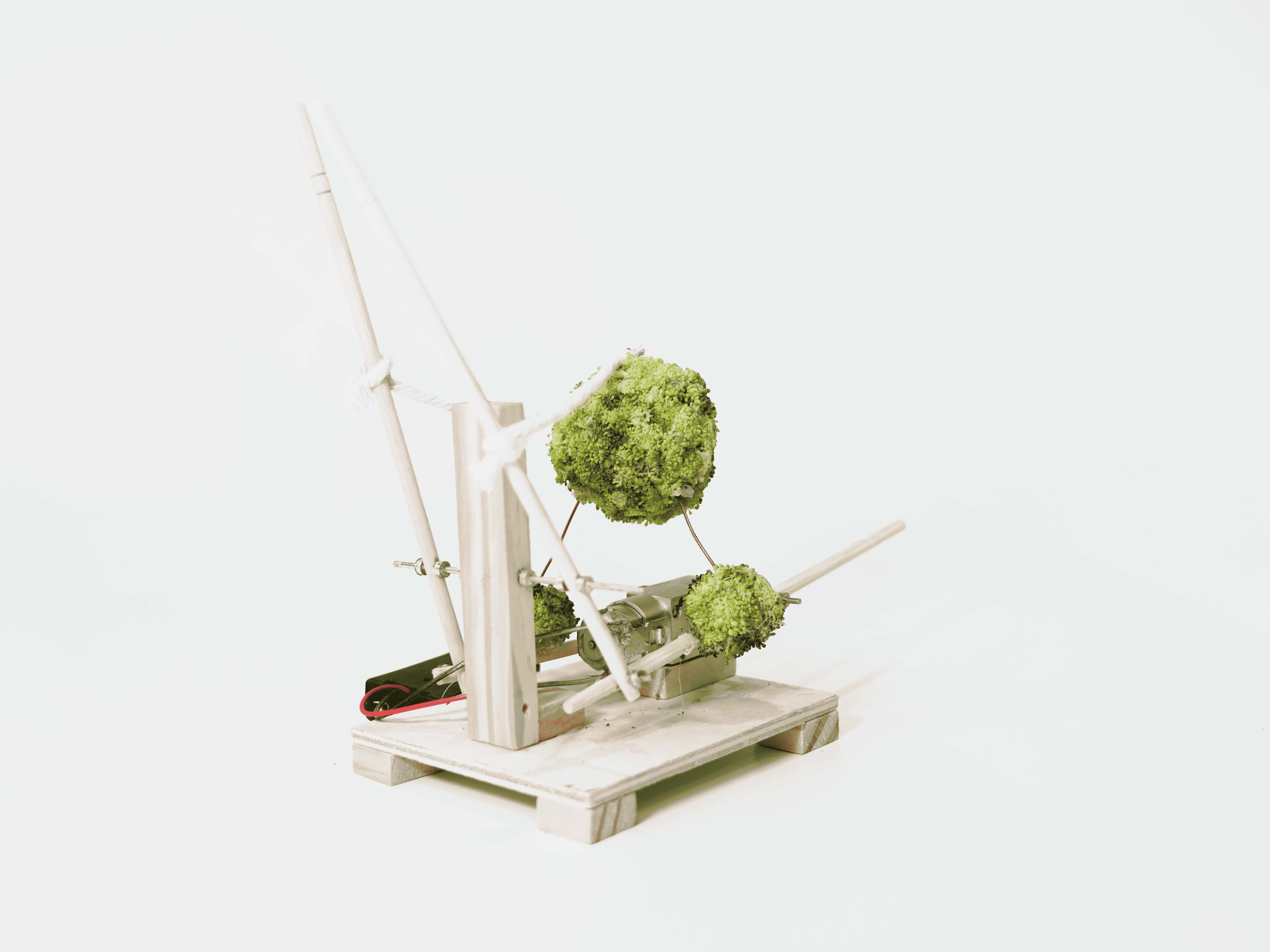}
    \caption{Cybroc: a series of kinetic art installations exploring populist longevity activism and transhumanism through the cyborgization of broccoli.}
    \label{fig:lab}
\end{figure}

\section{Introduction}

Populist longevity activism is a proliferating movement led by Silicon Valley's ultra-wealthy, focusing on promoting research, policies, and practices aimed at extending human lifespan and healthspan. For example, billionaire Bryan Johnson is dedicating significant efforts to this cause, promoting his ``Blueprint'' 
protocol\footnote{\url{https://protocol.bryanjohnson.com/}} to reverse aging.

\emph{Cybroc} is a series of kinetic art installations that features broccoli ``cyborgs''. Each broccoli is fitted with prosthetic limbs and mechanical attachments that enable it to perform longevity-enhancing exercises based on ``Blueprint'' protocols promoted by longevity activists. The exercises include cold plunges, treadmill running, arm-swinging (brachiation), and sled pushing—modern fitness routines that simulate primitive human survival activities.

The artwork's impact is heightened by its temporal nature. Over the course of the installation, the broccoli inevitably decays, despite (or perhaps because of) its intense ``exercise'' regimen. This deterioration serves as a metaphor for the limitations of biological systems and raises questions about the efficacy of extreme life extension efforts, the ethics of human enhancement beyond natural capabilities, and particularly transhumanist ideals.

Through its unique blend of humor, critique, and philosophical inquiry, \emph{Cybroc} challenges viewers to consider the boundaries between human and transhuman, the pursuit of immortality, and the potential consequences of pushing biological systems beyond their natural limits. It serves as a witty yet profound commentary on society's quest for eternal youth and the complex intersection of nature, technology, and human ambition.

\section{Related Works}

The longevity movement, championed by figures like Bryan Johnson with initiatives such as the ``Blueprint'' protocol, exemplifies contemporary aspirations to extend lifespan through technological intervention. Scholars have critiqued such efforts for their ethical and social implications, particularly the exacerbation of inequality. These critiques align with posthumanist perspectives, which challenge anthropocentric and technocentric narratives by emphasizing the ethical limits of human enhancement and the interconnectedness of all life forms. 

Transhumanism itself, as a philosophical and cultural framework, advocates human empowerment through technology. The key tenet of morphological freedom, which posits that humans should have the right to alter their bodies through technology, is central to this philosophy~\cite{inbook}. However, the transhumanist promise of extending life and enhancing human capabilities is critiqued for its potential to exacerbate social inequalities. Only a select few, often the wealthiest, can access such technologies, which may deepen the divide between privileged and marginalized communities. Furthermore, the ethical considerations surrounding human augmentation, particularly regarding the alteration of human nature, raise concerns about the potential consequences for individual identity and societal values.

Artistic endeavors that interrogate the interplay between the natural and the mechanical are not new. Works like \textit{Cyborg Botany} by Sareen and Maes \cite{10.1145/3290607.3313091} explore the integration of cybernetic systems into plant life to create new modes of interaction, emphasizing the blurred boundaries between the organic and the technological. Similarly, animistic projects like \textit{AniThings}~\cite{10.1145/2468356.2468746} present heterogeneous and symbolic narratives that blend functionality and subjectivity in everyday objects. These projects highlight the potential for hybrid entities to challenge conventional understandings of agency and functionality. The philosophical work of Bogdan Popoveniuc et al. \cite{10.1145/3532525.3532528} on extended reality provides a useful framework for understanding how technologies that enhance human capabilities might function within virtual and augmented environments. It questions whether XR-mediated enhancements, which blur the lines between reality and the virtual, ultimately enhance or distort our experience of the world and ourselves.

The philosophical questions raised by such projects align with critiques of transhumanism, as they confront the ethical and ecological implications of enhancing or manipulating living organisms.

\section{Artwork}
\begin{figure}
    \centering
    \includegraphics[width=1\linewidth]{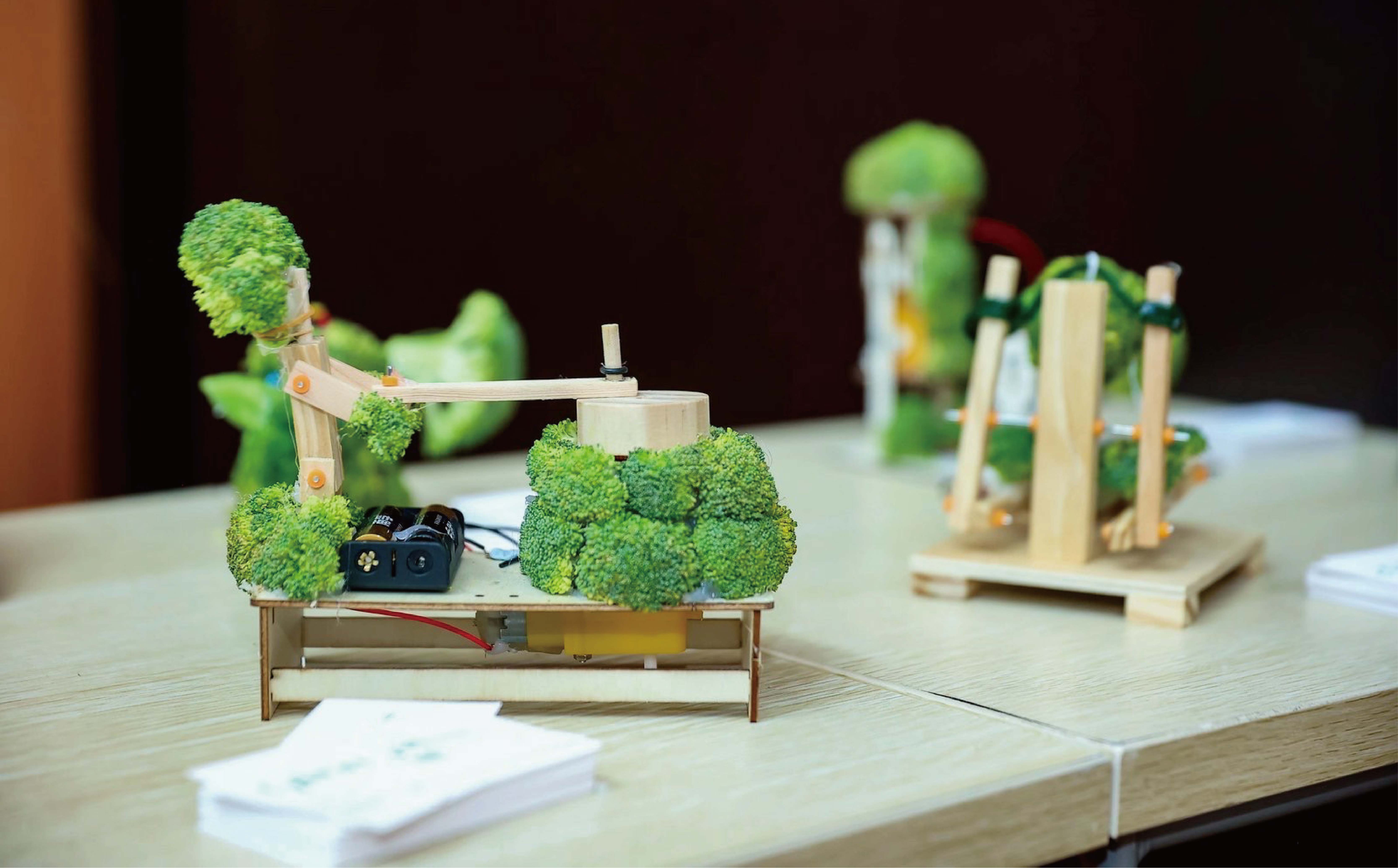}
    \caption{\emph{Cybroc} Installation}
    \label{fig:enter-label}
\end{figure}

\subsection{Artistic Concept}

The \emph{Cybroc} project uses broccoli—a symbol of health and longevity—as its central metaphor to critique contemporary longevity movements and transhumanist ideals. By augmenting broccoli with mechanical limbs to perform exaggerated fitness routines, the installation satirizes humanity's obsession with technological enhancement. These routines, such as treadmill running and arm-swinging, mimic primal survival tasks reframed as modern health practices, underscoring the absurdity of attempting to transcend biological limits through technology.

A key element of \emph{Cybroc} is the temporal nature of the artwork. As the broccoli inevitably decays despite its "enhanced" form, the project invites reflection on the futility of technological attempts to transcend the natural lifecycle. This decay serves as both a visual and symbolic critique, challenging the audience to consider the ethical boundaries of human enhancement.

\begin{figure}
\centering
\includegraphics[width=1\linewidth]{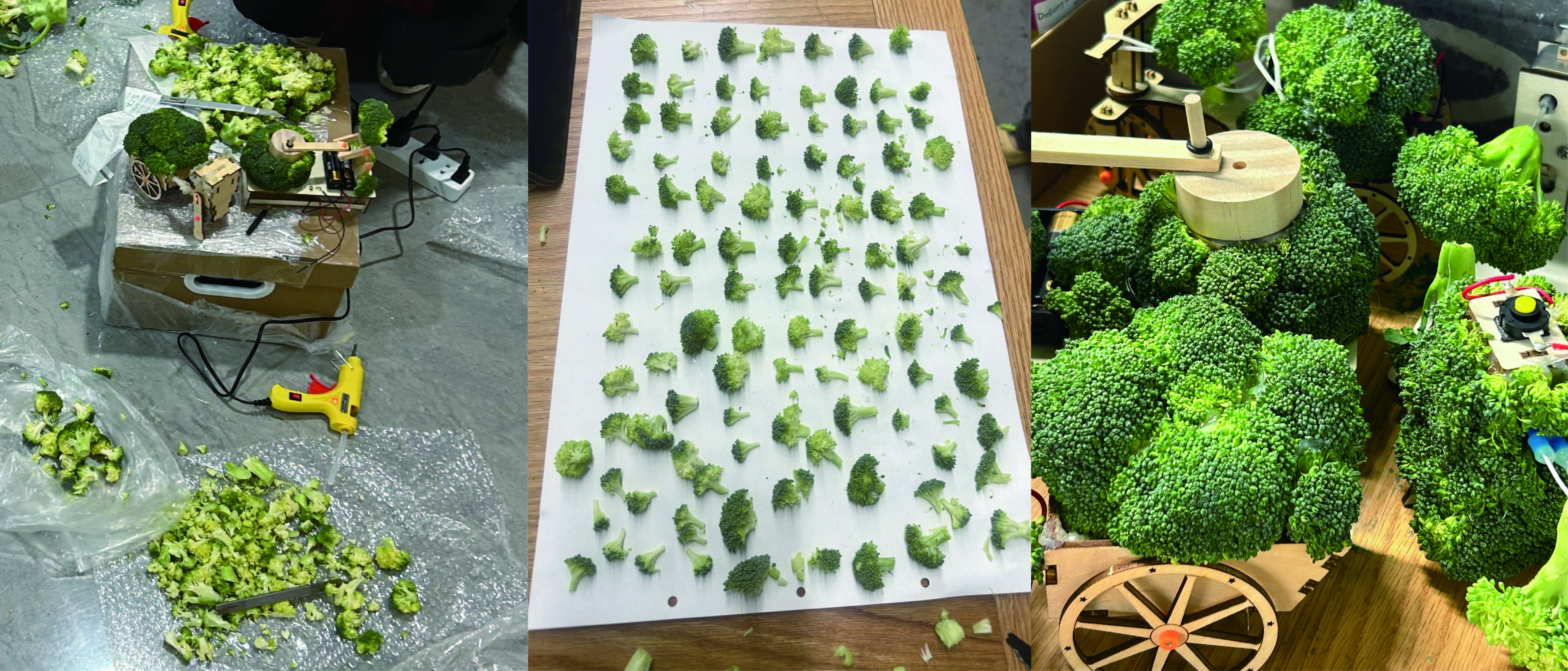}
\caption{Making \emph{Cybroc}: Working in progress}
\label{fig:wip}
\end{figure}

\paragraph{Materials}

Broccoli is chosen as the core material of the artwork for its rich symbolism. As a vegetable commonly associated with health and longevity, it becomes a powerful metaphor in the context of the longevity movement. By cyborgizing it with mechanical limbs, the artwork creates a visceral contrast between organic material and technological augmentation. The mechanical structures—comprising gears, chains, and metal supports—stand in stark contrast to the vegetable, intensifying the work's satirical tone.

\paragraph{Craftsmanship}
Each broccoli unit features a simple motor-driven mechanical system that enables repetitive motion. The design employs low-tech components such as basic gearboxes and belt drives to emphasize inefficiency as an artistic choice. Modular construction allows the devices to be easily disassembled and reassembled, facilitating portability and adaptability for different exhibition contexts. Its compact components ensure easy transportation, while the modular setup allows curators to customize the display to their venue.

\paragraph{``Inefficiency'' in Design}

The mechanical design is inherently inefficient, but this inefficiency is an artistic choice rather than a technical limitation. The simple, repetitive motions of the mechanical systems and the eventual decay of the broccoli imply that technological enhancement cannot negate the fragility of natural life. This inefficiency serves as a metaphor for the ultimate futility of efforts to transcend biological limitations.

\section{Exhibition and Propagation}

The \emph{Cybroc} installation is designed to engage audiences in a dynamic and reflective experience, blending humor, critique, and interaction. The exhibition strategy emphasizes accessibility, multisensory engagement, and meaningful dialogue between the artwork and its viewers.

\begin{figure}
    \centering
    \includegraphics[width=1\linewidth]{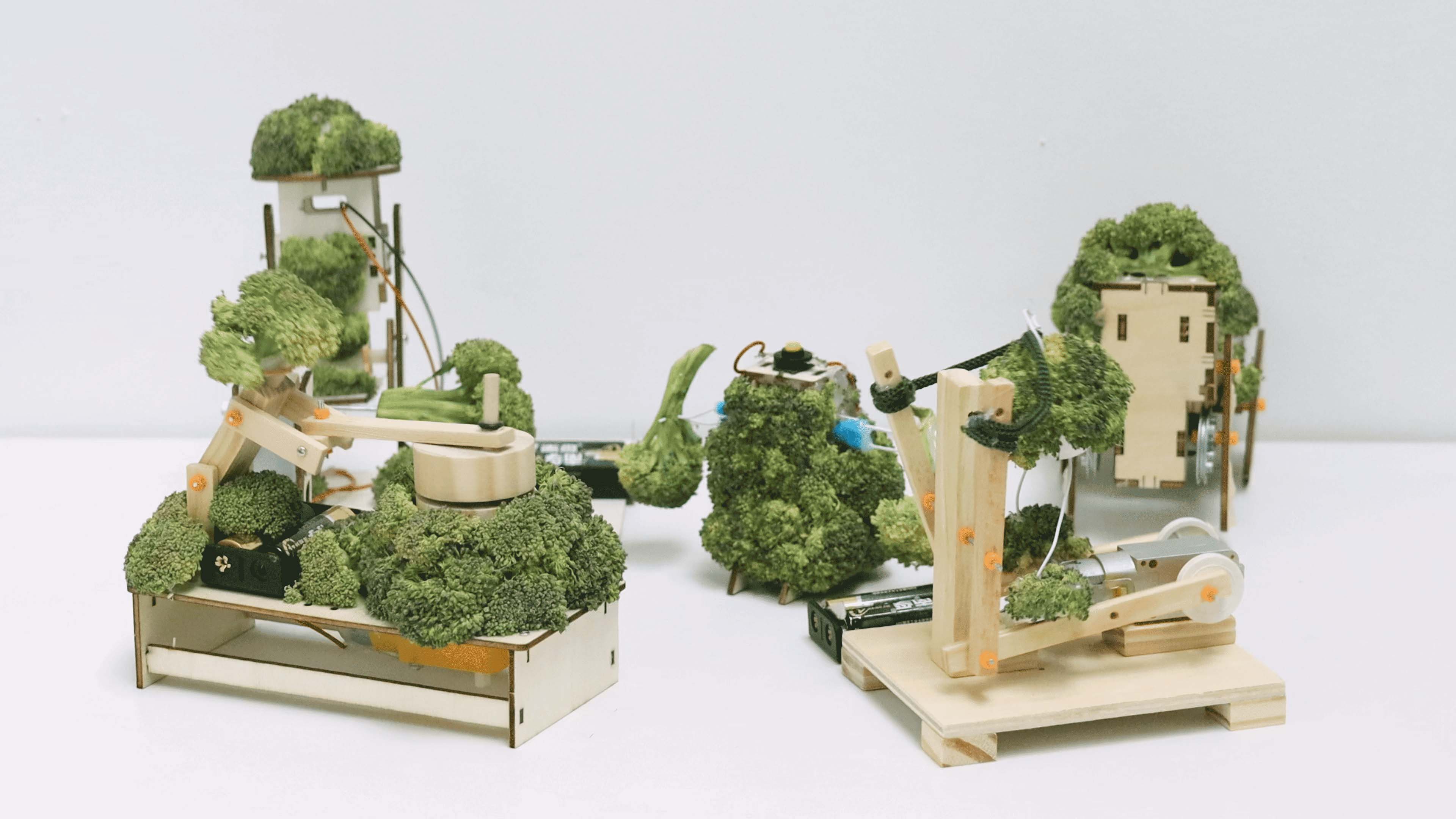}
    \caption{\emph{Cybroc} Family: A variety of robotic devices with different movement patterns. }
    \label{fig:family}
\end{figure}

\subsection{On-Site Creation}

\emph{Cybroc} installation features multiple mechanized broccoli units performing exaggerated fitness routines within a defined display area, as shown in Figure \ref{fig:family}. The curation and creation process is highly improvised onsite (Figure \ref{fig:wip}), based on available ready-made materials and the health and shape of purchased broccoli. For the exhibition, we typically create exhibits featuring cold plunges, treadmill running, brachiation (arm-swinging), and sled pushing.

\subsection{Audience Provocation}

\emph{Cybroc} is more than just the installation itself; it stimulates emotional engagement and intellectual interaction with the viewer. In dialogue with the artwork, the viewer is both observer and participant. Each installation demonstrates its assigned exercise task through mechanical movement, allowing the viewer to watch the broccoli perform repetitive physical movements through its cyborg augmentation, ultimately rotting and decaying. 

\begin{figure}
    \centering
    \includegraphics[width=1\linewidth]{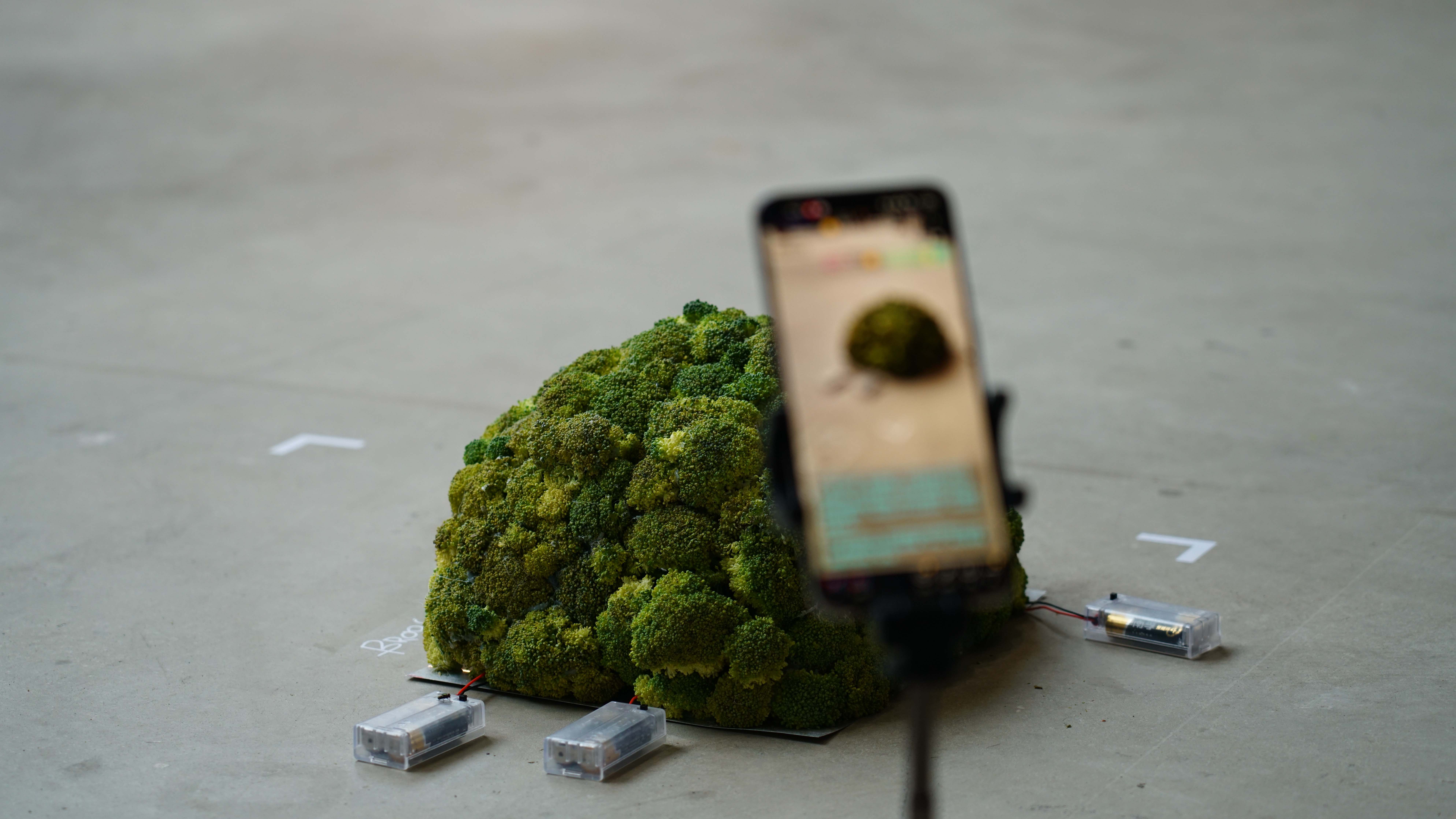}
    \caption{Live streaming for \emph{Cybroc} installation}
    \label{fig:live}
\end{figure}

\subsection{Live Streams and Short Videos}

The \emph{Cybroc} project leverages digital streaming platforms to extend its reach and foster engagement. Live streams of the exhibition allow remote audiences to experience the dynamic motion of the broccoli cyborgs in real time, as shown in Figure \ref{fig:live}.
Short videos tailored for social media platforms highlight the humor and critique embedded in the installation. These clips, featuring the absurdity of mechanical broccoli performing fitness routines, have garnered attention as shareable, meme-like content, furthering the project's visibility and relevance in online spaces.

\begin{figure}
    \centering
    \includegraphics[width=1\linewidth]{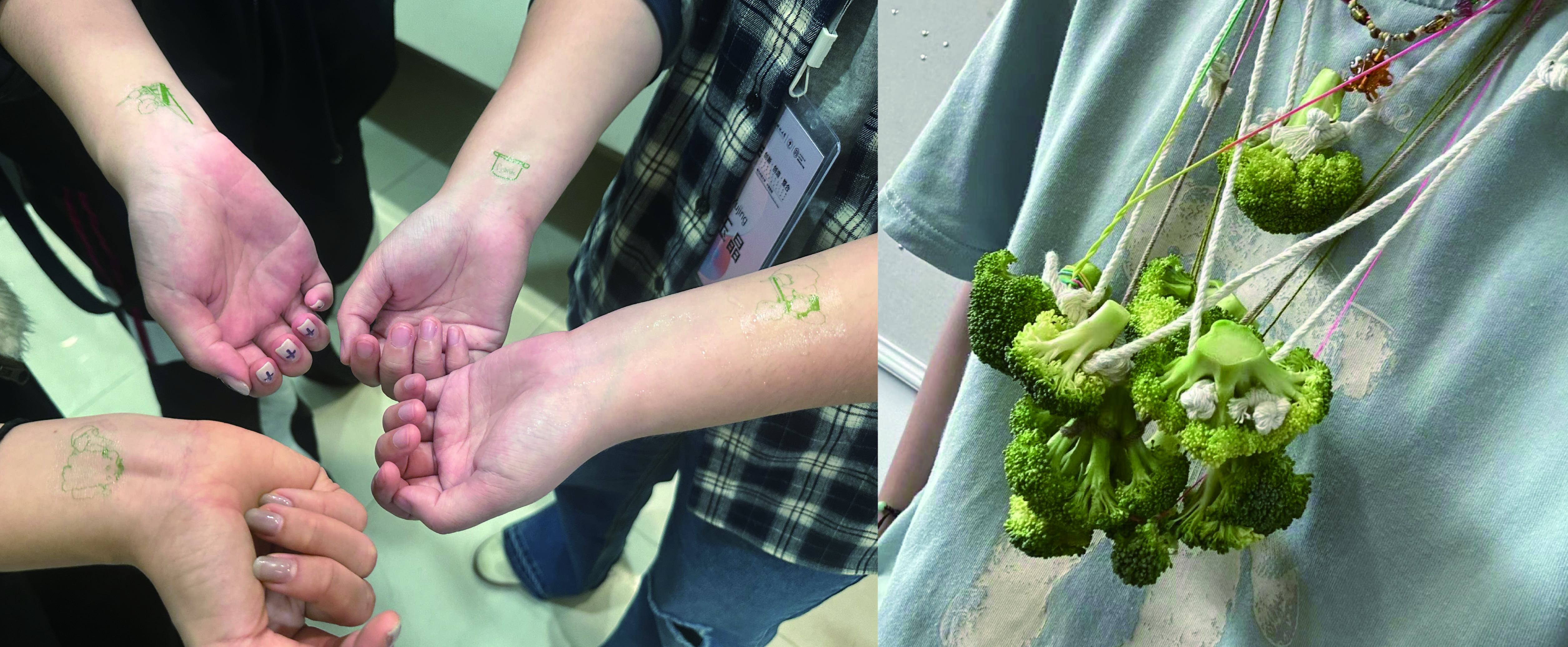}
    \caption{Wearable \emph{Cybroc} as Memorabilia}
    \label{fig:wearable}
\end{figure}

\subsection{Expanding Engagement with Physical Memorabilia}

To deepen its connection with audiences, \emph{Cybroc} introduced a line of physical memorabilia, including wearable broccoli and temporary \emph{Cybroc} tattoos featuring stylized broccoli cyborgs, as shown in Figure \ref{fig:wearable}. These tattoos became an instant hit at the exhibition, with visitors wearing them to showcase their engagement with the artwork.

This playful approach to engagement extends the project’s reach beyond the exhibition space. Attendees often share images of their tattoos on social media, inadvertently becoming ambassadors for the artwork.

\subsection{Meme-Based Propagation}
The visual humor of Cybroc naturally lends itself to meme culture. Images and GIFs of the mechanical broccoli in action have been shared widely on social platforms, resonating with audiences through their playful and ironic tone. The project’s critique of technological obsession is encapsulated in the absurdity of ``broccoli everywhere'', making it a perfect candidate for meme-driven virality. These memes encourage discourse on themes such as health, technology, and sustainability in a lighthearted yet thought-provoking manner.
\cite{10.1145/3449155}

\section{Discussion}

\subsection{Provocations for Questioning Longevity}

The experience provoked both humour and absurdity, while inviting viewers to consider the complex relationship between technology and biology. One participant noted that ``seeing broccoli rot while exercising made me question my obsession with fitness trackers''. The interactive aspect of this work lies in the reflection it encourages: does technology really help us transcend nature, or are we simply mechanizing our primal instincts? Through visual and emotional stimulation, \emph{Cybroc} challenges the viewer to question the validity of transhuman ideals and longevity activism.

\subsection{From Device Movement to Social Movement}
The exaggerated, mechanized movements of the \emph{Cybroc} device are a metaphor for the human obsession with bodily control. The work provokes public reflection on broader issues such as health inequalities and technological excess. \emph{Cybroc} acts as a catalyst for dialogue on redefining health and longevity, encouraging grassroots discussion of alternative, inclusive and sustainable approaches to well-being. The project also highlights gaps in access to life-extending technologies, raising questions about who benefits from these advances. By embedding these questions within an artwork, \emph{Cybroc} invites viewers to consider the ethical implications of prioritizing longevity over equity, and whether the social movement around extending life addresses our collective needs or merely individual desires.

\subsection{Rethinking the desire to transform nature}

The device criticizes the underlying ethos of modifying the natural body to prolong life and improve performance. Is it ethical, or even meaningful, to alter the processes of nature to such an extent? While technological advances promise to alleviate human suffering, the pursuit of immortality may lead to an overemphasis on individual longevity at the expense of social equity and ecological balance. \emph{Cybroc} asks the viewer to consider whether such a quest is intrinsically contradictory to the interconnectedness of all forms of life, and the inevitable cycles of decline and regeneration.

\subsection{Living for Joy or Longevity?}

The artwork also touches on the dilemma of whether one should strive for a strict, disciplined lifestyle, symbolized by 'eating healthy broccoli' to achieve longevity, or whether one should embrace a vibrant, imperfect life with unpredictability and happiness. This question resonates with broader social debates about health, happiness and meaning.

By presenting these tensions in a satirical and engaging framework, \emph{Cybroc} not only critiques the ethical boundaries of human progress, but also encourages reflection on the value of living. Through its funny and insightful commentary, the project emphasizes the importance of balancing technological progress with respect for natural and imperfect pleasures.

\section{Conclusion}

\emph{Cybroc} critiques society's obsession with technological solutions and the politics of the human body, exposing the underlying fear of aging and mortality. The pursuit of the "perfect body" through advanced technologies reflects a desire to defy nature's laws, but such ambitions exacerbate social inequalities by limiting access to these advancements.

In an age where the lines between human and machine are increasingly blurred, \emph{Cybroc} playfully challenges transhumanist ideals and raises urgent ethical questions about human enhancement. It explores the tension between humanity's natural limits and its relentless drive for perfection. Through the recontextualization of broccoli—a symbol of health—into a mechanized entity performing survival tasks, \emph{Cybroc} prompts reflection on whether the quest for longevity represents meaningful progress or a futile attempt to control nature.

By using humor and satire, \emph{Cybroc} delves into the complex relationship between nature, technology, and immortality, urging audiences to reconsider the ethical and societal implications of prioritizing technological progress over ecological balance and equity. The work contrasts the fragility of nature with the ambition of human enhancement, questioning whether the future lies in thoughtful evolution or in the mechanical imitation of a primal past.

\bibliographystyle{isea}
\bibliography{isea}

\end{document}